# Multiuser beam steering OWC system based on NOMA

Y. Zeng[1], Sanaa H. Mohamed[2], Ahmad Qidan[2], Taisir E. H. El-Gorashi[2], Jaafar M. H. Elmirghani[2]

[1]School of Electronic and Electrical Engineering, University of Leeds, LS2 9JT, United Kingdom
[2]Department of Engineering, King's College London, WC2R 2LS, United Kingdom
e-mail: ml16y5z@leeds.ac.uk.

**ABSTRACT**

In this paper, we propose applying Non-Orthogonal Multiple Access (NOMA) technology in a multiuser beam steering OWC system. We study the performance of the NOMA-based multiuser beam steering system in terms of the achievable rate and Bit Error Rate (BER). We investigate the impact of the power allocation factor of NOMA and the number of users in the room. The results show that the power allocation factor is a vital parameter in NOMA-based transmission that affects the performance of the network in terms of data rate and BER.

## 1. INTRODUCTION

Beam steering techniques were proposed in recent studies to address some of the limitations of Optical Wireless Communication (OWC) systems [1]-[3]. The transmitter of a beam steering OWC system provides a narrow beam that follows the mobile user. The narrow beam size reduces free space propagation loss and the Inter Symbol Interference (ISI) due to the reflection from walls and the ceiling [4]-[6]. LDs, which possesses high modulation bandwidth, can be efficiently integrated in beam steering OWC systems as the light source [6]-[9]. The authors in [10] used beam steering to achieve data rates of up to 25 Gbps. Beam steering techniques can be generally categorized into two groups. The first group is the mechanical beam steering (i.e., piezoelectric actuators-controlled lens), and the second group is non-mechanical beam steering (i.e., Spatial Light Modulator (SLM) and Micro Electromechanical Systems (MEMS) controlled mirrors). Compared with mechanical beam steering, non-mechanical beam steering consumes less power, have faster reaction speed, possess a smaller size and does not suffer from mechanical wear and failure [11]. Non-mechanical beam steering is the mainstream solution for beam steering OWC systems at present [1], [2], [12].

Fig. 1 shows the structure of a typical beam steering OWC transmitter. A digital Fresnel Zone Plate (FZP) pattern is rendered on the Spatial light modulator [1]. By changing the zone center of the FZP, the parallel light beams from the light source are focused on the far field with an arbitrary steering angle [13]-[15]. However, the maximum beam steering angle is limited by the pixel size of the SLM.

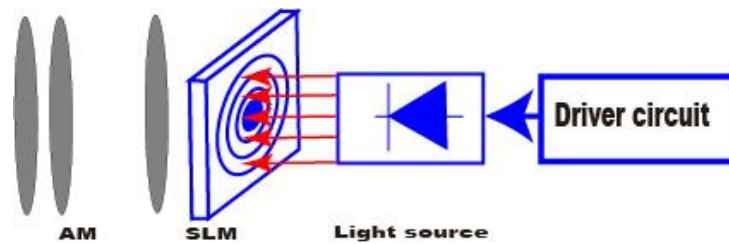

*Figure 1 The beam steering configuration of the transmitters.*

An Angle Magnifier (AM) constructed from 3 lens (shown in Fig.1) was proposed in [3] to extend the maximum beam steering angle of the transmitter. A beam steering transmitter with maximum beam steering angle of 30 degrees was reported in [1].

Most current research on beam steering OWC focused on systems with one user. The multiuser connectivity requirements are not considered. In this paper, we propose a multiuser beam steering OWC system based on Non-Orthogonal Multiple Access (NOMA) technology. The achievable date rate and Bit Error Rate (BER) of our proposed system are investigated using a simulation environment based on the ray tracing algorithm as proposed in [12] and [13].

The rest of the paper is organized as follows: The use of NOMA for multiuser beam steering OWC is introduced in Section 2. The simulation configuration and the results are presented in Section 3. The conclusions are presented in Section 4.

## 2. NOMA FOR OWC SYSTEMS

NOMA was proposed in [16] as a spectrum-efficient multiple access scheme for downlink in Visible Light Communication (VLC) systems. The signals of different users in NOMA systems are superimposed in the power domain with different power levels. The power levels are associated with users according to their channel conditions. Based on this, a user can share the entire frequency and time resources leading to increased spectral efficiency. NOMA allocates higher power levels to users with worse channel conditions than those with good channel conditions to balance the influence of the channels on different users. As a result, the user with the highest allocated power can directly decode its signal, while treating the signals of other users as noise. The rest of users in the system perform Successive Interference Cancellation (SIC) for the multi-signal separation for lower decoding orders prior to decoding their signals while treating the signal of high decoding orders as noise.

We consider a realistic indoor environment where multiple APs are mounted on the ceiling as shown in Fig. 2. All APs are connected with fiber and are connected to a central unit that controls the resources of the network and allocates them fairly among the users. Multiple users are randomly moving in the room. We assume that the APs find the location of the users based on the Location Estimation Algorithm (LEA) proposed in [17] and acquire perfect Channel State Information (CSI) of relevant users. Using NOMA and unipolar intensity modulation (LDs only works in positive area), each AP transmits real and positive signals to intended users. The signal from each AP is a superposition of the signals that convey the information for intended users with different power levels which is related to the CSI of different users. Let us now denote the set of users serviced by the $l^{th}$ AP by $U_l$. The optical peak power ($I_l$) and optical signal ($z_l$) transmitted from the $l^{th}$ AP can be given as [16]:

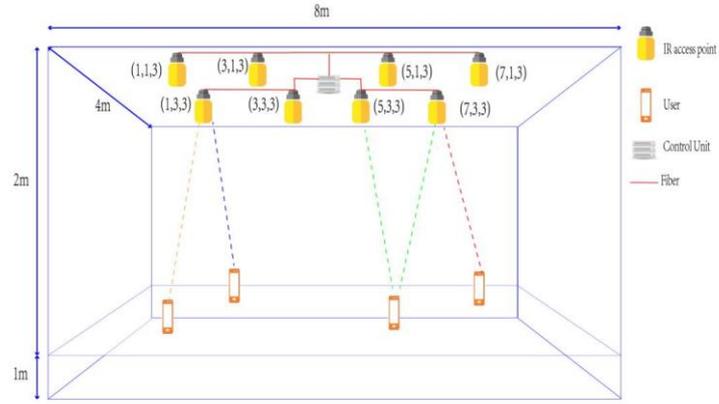

*Figure 2 Beam steering OWC system for multiuser.*

$$z_l = \sum_{j \in U_l} \eta \sqrt{P_{lj}} x_j, \quad I_l = \sum_{j \in U_l} \eta \sqrt{P_{lj}}, \qquad (1)$$

where $\eta$ is the quantum efficiency of the AP, $x_j$ is the signal symbol of user $j$, $z_l$ is the superposition optical signal from the AP, and $P_{lj}$ is the electrical power of $l^{th}$ AP allocated to user $j$. Considering there is a set of all users $S$ ($j \in S, all\ U_l \subseteq S$) staying in the room, the received signal at the receiver of user $j$ can be represented as [14]:

$$y_j = \sum_{l=1}^{L} RI_l z_l \otimes h_{lj} + n_j, \qquad j \in U_l, \qquad (2)$$

where $L$ is the total number of APs, $n_i$ is additive gaussian noise, which is sum of the contribution from the shot noise and thermal noise at the receiver of user $j$, and $h_{lj}$ is the impulse respond of the link. Using SIC, user $j$ decodes the signal of the lower decoding order and treat the signal of high decoding order as noise. The Signal to Interference and Noise Ratio (SINR) of user $j$ is given as [18]:

$$SINR_j = \sum_{l=1}^{L} \frac{(I_l R_j H_{lj})^2}{\sigma_j^2 + \sum_{k>j}(I_{lk} R_j H_{lj})^2} \qquad (3)$$

where $H_{lj}$ is the channel gain of the channel between $l^{th}$ AP and user $j$ and $\sigma_j^2$ is the additive Gaussian noise of user $j$ due to ambient light, pre-amplifier and signal related noise. The achievable rate of user $j$ is given as [15]:

$$r_j = \frac{1}{2} B \log_2(1 + SINR_j), \qquad (4)$$

where the scaling factor ($\frac{1}{2}$) is due to the Hermitian symmetry and $B$ is the bandwidth. The decoding order of the users is related to their CSI, $h_{lj}$, which is collected during the process of the AP locating the user. After APs report CSIs of all users to the central unit, the decoding order of user $j$ is decided by the sum of CSI, $h_{lj}$, relating to all APs. The decoding priorities of users are decided in reverse order based on the CSIs of all users. We consider user $j$ and $j-1$ ($j, j-1 \in U_l$) to be adjacent in the decoding sequence and the decoding priority of user $j-1$ is higher than user $j$. For $l^{th}$ AP, the power allocation of the $j$ and $j-1$ sorted users are revealed as:

$$P_{j-1} = \alpha P_j \qquad (5)$$

where $\alpha$ is the power allocation factor.

## 3. SIMULATION SETUP AND RESULTS

To evaluate the advantages of the proposed method, a simulation was performed considering an empty room with dimensions of 8 m × 4 m × 3 m (length × width × height). The plaster walls are assumed to reflect light rays in a form close to a Lambertian function. Therefore, the walls, ceiling and floor were modelled as Lambertian reflectors with reflectivity of 80% and 30%, respectively [17]. Reflections from doors and windows are identical to reflections from walls. The transmitted signals are reflected by the room reflecting surfaces. The surfaces are divided into a number of equally-sized, square-shaped reflection elements. The reflection elements have been treated as small transmitters that diffuse the received signals from their centers in a Lambertian pattern [19] [20]. It is noted that third-order and higher reflections do not produce a significant change in the received optical power, and therefore reflections up to second order only are considered [5]-[7]. In this work, the surface element size is set to 5cm × 5cm for the first-order reflections and 20cm × 20cm for the second-order reflections. The illumination system is constructed using 8 LDs as in [10].

The locations of the APs are shown as Fig. 2. Considering 4 to 10 users in the room, the locations of them are randomly generated following a uniform distribution. For each scenario (i.e., different number of users), the simulation is repeated for 100 times to generate different user locations, while the power allocation factor $\alpha$ is varied from 0.2 to 0.8. The rest of the parameters are shown in Table 1.

*Table 1. Simulation parameters*

| Parameter | Configuration | |
|---|---|---|
| Room setup | | |
| Length, Width, Height | 8m, 4m, 3m | |
| Reflectivity of walls | 0.8 | |
| Reflectivity of ceiling | 0.8 | |
| Reflectivity of floor | 0.3 | |
| IR AP | | |
| Quantity | 8 | |
| Location | (1,1,3) (1,3,3) (1,5,3) (1,7,3) (3,1,3) (3,3,3) (3,5,3) (3,7,3) | |
| The average optical power | 1mW | |
| beam divergency | 2.1mrad | |
| Users | | |
| Quantity | 4 to 10 | |
| Elevation | 90º | |
| Azimuth | 0º | |
| Area | 1cm$^2$ | |
| The responsivity of IR | 0.5A/W | |
| The responsivity of illumination | 0.4A/W | |
| FOV | 90º | |
| Resolution | | |
| Time bin duration | 0.01 ns | |
| Bounces | 1 | 2 |
| Surface elements | 32000 | 2000 |
| Wavelength | 850nm | |
| Bandwidth (BW) | 10 GHz | |

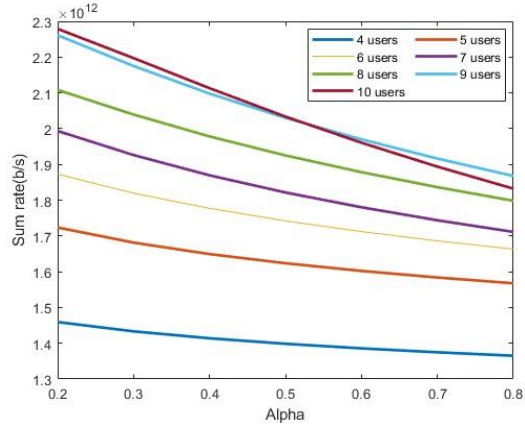

*Figure 3 The sum rates of users versus the power allocation factor of NOMA.*

Fig. 3 shows the sum rate of the system versus the power allocation factor, considering different numbers of users. It can be seen that the sum rate decreases as the power allocation factor increases. Note that, an increase in $\alpha$ means that the users classified as weak users according to their channel gains receive low power, and therefore low SINR is experienced by these users. On the other hand, the strong users receive high power as $\alpha$ increases. However, the sum rate decreases due to the fact that the strong users apply SIC to decode their information, and high power allocated to the strong users means noise enhancement. The figure further shows that the sum rate increases with the number of users, which is expected as all the users in the network experience different channel gains.

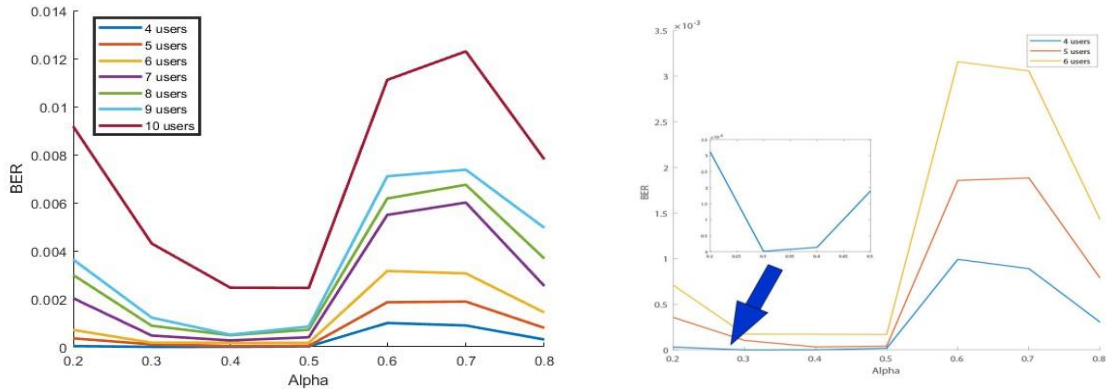

*Figure 4 Average BER against the power allocation factor*

Fig. 4(a) shows the average BER for different number of users against $\alpha$, and Fig. 4(b) zooms up the BER of the scenarios where 4, 5, and 6 users are considered. For $\alpha \leq 0.5$, the BER of all scenarios decreases as the power allocation factor increases, since each user receives a decent power level to decode the desired information. However, when $\alpha > 0.5$, high power is allocated to the users with high channel gains, which negatively impacts the SINR of the system. Note that, the signal intensity difference between the users increases when $\alpha = 0.8$, which leads to high probability of successive decoding for the strong users at the cost of low fairness. However, the results show that a power allocation factor of 0.4 is a good choice to acquire high sum rate and low BER.

## 4.CONCLUSION

In this paper, a beam steering OWC multi-user system based on NOMA is introduced. We investigated the achievable bit rate and BER of the proposed system while changing the power allocation factor, $\alpha$. The achievable rate of users is degraded when increasing the power allocation factor while the BER fluctuates with the increase of the power allocation factor. The results show that a balance between the BER and the achievable rate can be achieved when $\alpha = 0.4$.

**ACKNOWLEDGMENTS**

The authors would like to acknowledge funding from the Engineering and Physical Sciences Research Council (EPSRC) TOWS (EP/S016570/1) project.